\documentclass[
superscriptaddress,
preprint,
showpacs,
amsmath,amssymb,
floatfix,
lengthcheck,
]{revtex4-1}

\usepackage{natbib}
\usepackage[dvipdfmx]{graphicx,color}
\usepackage{dcolumn}
\usepackage{bm}
\usepackage{ulem} 

\newcommand{\uar}{\uparrow}
\newcommand{\dar}{\downarrow}
\newcommand{\lag}{\langle}
\newcommand{\rag}{\rangle}
\newcommand{\lt}{\left}
\newcommand{\rt}{\right}

\newcommand{\hspc}{\hspace{1em}}

\newcommand{\mrm}{\mathrm}
\newcommand{\mbf}{\mathbf}
\newcommand{\mcl}{\mathcal}
\newcommand{\mbb}{\mathbb}

\newcommand{\der}{\mrm{d}}
\newcommand{\pder}{\mrm{\partial}}

\newcommand{\im}{\mrm{i}}
\newcommand{\ex}{\mrm{e}}

\begin{document}

\title{Reentrant Topological Phase Transition in a Bridging Model\\ between Kitaev and Haldane Chains}

\author{Takanori Sugimoto}
\email[Electronic address: ]{sugimoto.takanori@rs.tus.ac.jp}
\affiliation{Department of Applied Physics, Tokyo University of Science, Katsushika, Tokyo 125-8585, Japan}
\author{Mitsuyoshi Ohtsu}
\altaffiliation[Present address: ]{Department of Basic Science, University of Tokyo, Meguro, Tokyo 153-8902, Japan}
\affiliation{Department of Applied Physics, Tokyo University of Science, Katsushika, Tokyo 125-8585, Japan}
\author{Takami Tohyama}
\affiliation{Department of Applied Physics, Tokyo University of Science, Katsushika, Tokyo 125-8585, Japan}

\date{\today}

\begin{abstract}
We present a reentrant phase transition in a bridging model between two different topological models: Kitaev and Haldane chains.
This model is activated by introducing a bond alternation into the Kitaev chain [A. Yu Kitaev, Phys.-Usp. {\bf 44} 131 (2001)].
Without the bond alternation, finite pairing potential induces a topological state defined by zero-energy Majorana edge mode, while finite bond alternation without the pairing potential makes a different topological state similar to the Haldane state, which is defined by local Berry phase in the bulk.
The topologically-ordered state corresponds to the Su-Schrieffer-Heeger state, which is classified as the same symmetry class.
We thus find a phase transition between the two topological phases with a reentrant phenomenon, and extend the phase diagram in the plane of the pairing potential and the bond alternation by using three techniques: recursive equation, fidelity, and Pfaffian.
In addition, we find that the phase transition is characterized by both the change of the position of Majorana zero-energy modes from one edge to the other edge and the emergence of a string order in the bulk, and that the reentrance is based on a sublattice U(1) rotation.
Consequently, our study and model do not only open a direct way to discuss the bulk and edge topologies, but demonstrate an example of the reentrant topologies.
\end{abstract}


\maketitle

\section{Introduction}
The Kitaev chain~\cite{Kitaev01} has been attracted much attention because of a topological aspect associated with Majorana fermions~\cite{Wilczek09} in condensed matters.
In this model, every real-space fermion is transformed into Majorana fermions, and the Hamiltonian can be rewritten by a canonical form constructed by paired Majorana fermions. 
Since a pair of Majorana fermions corresponds to a fermionic number operator, a coupling constant plays the role of chemical potential of the fermions.
Kitaev has indicated existence of unpaird Majorana fermions on the edges of an infinite-length chain~\cite{Kitaev01}, that is, an emergent zero-energy mode called Majorana zero mode (MZM), which changes the fermionic parity.
The MZM appears if the Majorana number defined by $\pm 1$ has the non-trivial value ($-1$), and thus the phase with MZM is regarded as a Z$_2$ topological phase defined by the Majorana number.

This model is also obtained by the Jordan-Wigner transformation of an XY-type spin-$\frac{1}{2}$ chain with non-zero XY anisotropy.
The ground-state degeneracy of MZM corresponds to that of N\'eel states, $|\uar\dar\uar\cdots\dar\rag$ and $|\dar\uar\dar\cdots\uar\rag$. 
Another topology has been reported by Hatsugai~\cite{Hatsugai07} in Heisenberg spin chain with a bond alternation but no anisotropy.
This model is so-called the spin-Peierls model~\cite{Chesnut66,Pincus71,Pytte74,Cross79,Nakano80}, which also corresponds to the Su-Schrieffer-Heeger (SSH) model~\cite{Su79,Takayama80,Sugimoto12}.
The preceding study~\cite{Hatsugai07} has shown that an emergent alternating Z$_2$ topological order defined by local Berry phase in a valance-bond solid is regarded as a dimer-singlet ground state.
In addition, this phase is smoothly connected to that in a spin chain whose nearest-neighbor interaction altenates between ferromagnetic and anti-ferromagnetic.
Since the ground state of the alternating spin chain is equal to the Haldane state of the $S=1$ one-dimensional spin system~\cite{Haldane83}, the Z$_2$ topological order originates from a string order emerging in the Haldane chain~\cite{Affleck88,Nijs89,Tasaki91,Kennedy92}. 
Another recent study on the Haldane chain also reported that this $S=$(an odd integer) Haldane chain is classified as a symmetry-protected topological phase~\cite{Pollmann12}, where the Majorana number is trivial, $\mcl{M}=1$.
Therefore, the bond alternation can activate a Z$_2$ topologically-ordered phase in the Kitaev chain, and a phase transition between the two different Z$_2$ topological phases is expected.
In this paper, we thus investigate effects of the bond alternation on the topological phases.

Recently, some pioneering works on this effects have shown the phase diagram with a topological transition~\cite{Wakatsuki14,Xiong15,Zhou16,Bahari16}.
In these studies, an SSH-type ground state has been reported with a determination of symmetry class, which is the same as the topological state with an MZM.
However, these studies have not mentioned the topological properties of bulk, and thus, in this paper, we make it clear with an extended phase diagram obtained by alternative approaches.

The contents of this paper are as follows.
In Sec.~II, we present the spinless-fermion Hamiltonian of Kitaev chain with bond alternation under open boundary condition.
The Hamiltonian of Majorana-fermions and $S=\frac{1}{2}$ spins are also obtained by exact transformations.
Additionally, we mention two limits of this model: finite pairing potential without bond alternation and finite bond alternation without pairing potential, where the Majorana number is regarded as a topological invariant.
The section III is devoted to system-size parity and phase diagram defined by the Majorana number in the plane of the pairing potential and the bond alternation.
Phase boundaries are determined by three techniques: recursive equation, fidelity, and Pfaffian of the Majorana fermions.
In these calculations, we obtain the phase boundaries consistent with a change of the Majorana number caused by switching the position of Majorana edge modes.  
Furthermore, we discuss dispersion relation with relation to a winding number of the spinless fermions, and show a change of string order though the topological transition numerically obtained by variational matrix-product state (MPS) method in Sec.~IV.
In the dispersion relation, we find a characteristic difference between two phase boundaries. 
In Sec.~V, summary is given with a comment on potential application of our system.

\section{Model}
We consider the Hamiltonian of an $N$-site Kitaev chain with bond alternation given by,

\begin{equation}
\mcl{H}_{\mrm{bulk}}=t\sum_{j=1}^{N-1} (1-\gamma\ex^{\im\pi j}) (c_j^\dag c_{j+1}+\lambda c_j^\dag c_{j+1}^\dag+\mrm{H.c.}),
\label{eq:ferham}
\end{equation}
where $\gamma$ and $\lambda$ are bond alternation and normalized pairing interaction, respectively.
The creation and annihilation operators of $j$th-site spinless fermion are represented by $c_j^\dag$ and $c_j$.
We take the open boundary condition for the chain.

The Hamiltonian (\ref{eq:ferham}) can be exactly mapped into two models: a Majorana fermion model and a spin-$\frac{1}{2}$ model.
The former is obtained by the introduction of Majorana fermions, $a_j=c_j^\dag+c_j$ and $b_j=\im(c_j^\dag-c_j)$, into the Hamitonian $\mcl{H}_{\mrm{bulk}}$:
\begin{equation}
\mcl{H}_{\mrm{bulk}}=\frac{\im t}{2}\sum_{j=1}^{N-1} (1-\gamma\ex^{\im\pi j}) \lt[(1-\lambda)a_jb_{j+1}-(1+\lambda)b_j a_{j+1}\rt],
\label{eq:majoham}
\end{equation}
where the Majorana fermions have anti-commutation relation, $\{a_i,a_j\}=\{b_i,b_j\}=2\delta_{i,j}$ and $\{a_i,b_j\}=0$.
The Jordan-Wigner transformation, $c_j=\ex^{\im \phi_j} \sigma_j^-$, and its Hermite conjugate give the latter model:
\begin{equation}
\mcl{H}_{\mrm{bulk}}=\frac{t}{2}\sum_{j=1}^{N-1}(1-\gamma\ex^{\im\pi j}) \lt[(1+\lambda)\sigma_j^x \sigma_{j+1}^x+(1-\lambda)\sigma_j^y \sigma_{j+1}^y\rt],
\label{eq:spiham}
\end{equation}
with $\phi_j=\pi\prod_{i(<j)} \lt[\frac{1}{2}\lt(\sigma_i^z+1\rt)\rt]$.
Here, we use the Pauli matrices $\bm{\sigma}$ and the ladder operators $\sigma^{\pm}=\frac{1}{2}\lt(\sigma^x\pm\im\sigma^y\rt)$.

To comfirm the presence of MZM, we rewrite the Majorana Hamiltonian (\ref{eq:majoham}) as the following canonical form:
\begin{equation}
\mcl{H}_{\mrm{bulk}}=\frac{\im t}{2}\sum_k \epsilon_k a_k^\prime b_k^\prime
\label{eq:kanoham}
\end{equation}
with $a_k^\prime=\sum_j u_{k,j} a_j$ and $b_k^\prime=\sum_j v_{k,j} b_j$ obtained by the orthogonal transformation of Majorana fermions. Both $a_k^\prime$ and $b_k^\prime$ satisfy the Majorana-type anti-commutation relation.
Equation (\ref{eq:kanoham}) is noting but the singular-value decomposition of the matrix form $\mbf{H}_c$ for the Majorana Hamiltonian (\ref{eq:majoham}) with respect to the vectors $\bm{a}=(a_1,a_2,\cdots,a_N)$ and $\bm{b}=(b_1,b_2,\cdots,b_N)$:
\begin{equation}
\mcl{H}_{\mrm{bulk}}=\frac{\im t}{2} \bm{a}\, \mbf{H}_c\, \bm{b}^{\mrm{T}} = \frac{\im t}{2} \bm{a}\, \mbf{U}^{\mrm{T}} \mbf{\Upsilon}_c \mbf{V}\, \bm{b}^{\mrm{T}},
\end{equation}
where $\mbf{\Upsilon}_c$ is a diagonal matrix with singular values $(\mbf{\Upsilon}_c)_{k,k^\prime}=\epsilon_k\delta_{k,k^\prime}$, and the special orthogonal matrices correspond to $(\mbf{U})_{k,j}=u_{k,j}$ and $(\mbf{V})_{k,j}=v_{k,j}$ with $\mrm{det}[\mbf{U}]=\mrm{det}[\mbf{V}]=1$.
If there is a mode $\mrm{k}_0$ satisfying $\epsilon_{\mrm{k}_0}=0$, Majorana fermions $a_{\mrm{k}_0}^\prime$ and $b_{\mrm{k}_0}^\prime$ commute with the Hamiltonian, $[a_{\mrm{k}_0}^\prime,\mcl{H}_{\mrm{bulk}}]=[b_{\mrm{k}_0}^\prime,\mcl{H}_{\mrm{bulk}}]=0$. 
Combining this fact with the following relation 
\begin{equation}
\frac{\im}{2} a_{\mrm{k}_0}^\prime b_{\mrm{k}_0}^\prime = d_{\mrm{k}_0}^\dag d_{\mrm{k}_0} - \frac{1}{2},
\end{equation}
corresponding to the number operator of the fermion defined by $d_{\mrm{k}_0}=a_{\mrm{k}_0}^\prime+\im b_{\mrm{k}_0}^\prime$, we can say that there is a zero-energy fermion constructed by the two Majorana operators $a_{\mrm{k}_0}^\prime$ and $b_{\mrm{k}_0}^\prime$. This is thus an MZM.

At $\gamma=0$, the condition for the MZM with $\epsilon_{\mrm{k}_0}=0$ corresponds to the following recursive equations:
\begin{equation}
u_{\mrm{k}_0,j+2} = -\Lambda u_{\mrm{k}_0,j} \hspc \mrm{and} \hspc   v_{\mrm{k}_0,j+2} = -\Lambda^{-1} v_{\mrm{k}_0,j} \label{eq:rec}
\end{equation}
with $\Lambda=(1-\lambda)/(1+\lambda)$ and boundary constraints
\begin{equation}
u_{\mrm{k}_0,2}=u_{\mrm{k}_0,N-1}=0 \hspc \mrm{and} \hspc v_{\mrm{k}_0,2} = v_{\mrm{k}_0,N-1}=0. \label{eq:bcon}
\end{equation}
In the $N=2n+1$ system ($n\in \mbb{N}$), there are always solutions such as $u_{\mrm{k}_0,2i}=0$ and $u_{\mrm{k}_0,2i+1}=\lt(-\Lambda\rt)^i u_{\mrm{k}_0,1}$ ($i=1,2,\cdots,n$) with $u_{\mrm{k}_0,1}=\sqrt{(1-\Lambda^2)/(1-\Lambda^{2n+2})}$, leading to $[a_{\mrm{k}_0}^\prime,\mcl{H}_{\mrm{bulk}}]=0$.
The presence of the solution is a consequence of the Kramer's doublet due to half-integer magnetization in the odd-number $N$-site spin system.

If $N$ is even and finite number, there is no solution for MZM except for $\lambda=\pm1$. 
For $\Lambda\neq 1$, however, a coupling energy exponentially decreases with increasing the system size, leading to $\epsilon_{\mrm{k}_0}=0$.
Therefore, we find solutions for MZM in the thermodynamical limit $N\rightarrow\infty$ keeping $N$ even.
In this case, the Majorana fermion $a_{\mrm{k}_0}^\prime$ localizes at one edge and $b_{\mrm{k}_0}^\prime$ localizes at the other edge, because their amplitudes $u_{\mrm{k}_0,j}$ decreases if $v_{\mrm{k}_0,j}$ increases with increasing $j$, and vice versa.
This MZM constructed by the Majorana fermions localizing at the edges is important for exhibiting a non-trivial topological number, i.e., Majorana number $\mcl{M}=-1$ defined by Pfaffian.
Hereafter, we call MZM constructed by $a_{\mrm{k}_0}^\prime$ ($b_{\mrm{k}_0}^\prime$) the $a$-type ($b$-type) MZM and the phase characterized by $\mcl{M}=-1$ the MZM phase.

Before discussing topological transition in our Hamiltonian, we explain the case for $\lambda=0$, that is, a bond-alternating isotropic XY spin chain. 
Finite bond alternation $0<\gamma\leq 1$ gives rise to valance-bond solid in the ground state, where singlet is locally constructed on the bonds with larger exchange interaction $(1+\gamma)t/2$.
This phase is smoothly connected to that in the region of $\gamma>1$, where the ground state is similar to the Haldane state of the $S=1$ one-dimensional spin system, because neighboring bonds alternate between ferromagnetic and antiferromagnetic ones. 
This state has topological order defined by local Berry phase, known as symmetry-protected topological phase~\cite{Pollmann12}, where the Majorana number is trivial, $\mcl{M}=1$. 
Therefore, our model has potential for phase transition between the MZM phase and the Z$_2$ topologically ordered phase similar to the Haldane state. 

\section{Topological Properties of Majorana Fermions}

In this section, we discuss phase transition between the MZM phase and the Z$_2$ topologically ordered phase.
We use three techniques: recursive equation to obtain a solution of the MZM, fidelity around the critical point, and Pfaffian for the twisted boundary condition in each phase.

\subsection{Recursive equation}
We start with the recursive-equation analysis on the topological transition~\cite{Fendley12}.
Similar to the case of $\gamma=0$, the recursive equations for MZM is obtained by the singular-value decomposition as follows,
\begin{equation}
u_{\mrm{k}_0,j+2} = -\Lambda\Gamma_j u_{\mrm{k}_0,j} \hspc \mrm{and} \hspc v_{\mrm{k}_0,j+2} = -\Lambda^{-1}\Gamma_j v_{\mrm{k}_0,j}\ ,
\label{eq:uv}
\end{equation}
where $\Gamma_j=(1-\gamma\ex^{\im\pi j})/(1+\gamma\ex^{\im\pi j})$ and boundary constraints are the same as Eq.~(\ref{eq:bcon}).
There are two possible modes for each type of MZMs, i.e., an even-site mode where $u_{\mrm{k}_0,2i-1}=0$ for the $a$-type MZM ($v_{\mrm{k}_0,2i-1}=0$ for the $b$-type MZM), and an odd-site mode where $u_{\mrm{k}_0,2i}=0$ for the $a$-type MZM ($v_{\mrm{k}_0,2i}=0$ for the $b$-type MZM) with $i=1,2,\cdots<(N-1)/2$.

If $N$ is even, we impose a fixed-end boundary constraint on the left (right) edge for the even-site (the odd-site) mode.
In this case, the equations in (\ref{eq:uv}) give the following four cases, provided that neither $|\Lambda|=|\Gamma_1|$ nor $|\Lambda|=|\Gamma_1|^{-1}$.
\begin{enumerate}
\item $|\Lambda|<\min\{|\Gamma_1|,|\Gamma_1|^{-1}\}$\\
Since $|\Lambda\Gamma_1|<1$, we find two solutions in the thermodynamical limit $N\to\infty$: the odd-site mode with $u_{\mrm{k}_0,2i+1}=(-\Lambda\Gamma_1)^{i}u_{\mrm{k}_0,1}$ and the even-site mode with $v_{\mrm{k}_0,N-2i}=(-\Lambda\Gamma_1)^{i}v_{\mrm{k}_0,N}$.
Thus, the $a$-type and $b$-type MZMs localize on the left and right edges, respectively. 

\item $|\Gamma_1|<|\Lambda|<|\Gamma_1|^{-1}$\\
The number of solutions in this case is not two but four. 
The boundary constraints allow both even-site and odd-site modes for each type of MZMs.
Therefore, there are both the $a$-type and $b$-type MZMs on both edges. 

\item $|\Gamma_1|^{-1}<|\Lambda|<|\Gamma_1|$\\
We cannot find any solution in this case. 

\item $|\Lambda|>\max\{|\Gamma_1|,|\Gamma_1|^{-1}\}$\\
This case is opposite to the case 1.
Since $|\Lambda^{-1}\Gamma_1|<1$, solutions in the thermodynamical limit appear as the even-site mode with $u_{\mrm{k}_0,N-2i}=(-\Lambda^{-1}\Gamma_1)^{i}u_{\mrm{k}_0,N}$ and the odd-site mode with  $v_{\mrm{k}_0,2i+1}=(-\Lambda^{-1}\Gamma_1)^{i}v_{\mrm{k}_0,1}$
Thus, the $a$-type and $b$-type MZMs appear on the right and left edges, respectively. 
\end{enumerate}

The conditions for $|\Lambda|$ and $|\Gamma_1|$ are determined by the parameters $\lambda$ and $\gamma$, since $\Lambda=(1-\lambda)/(1+\lambda)$ and  $\Gamma_1=(1+\gamma)/(1-\gamma)$.
For example, the case 1  corresponds to the condition $\lambda\in(\lambda_{\mrm{c1}},\lambda_{\mrm{c2}})$ with $\lambda_{c1}=\min\{|\gamma|,|\gamma|^{-1}\}$ and $\lambda_{c2}=\max\{|\gamma|,|\gamma|^{-1}\}$.

\begin{figure}[t]
\includegraphics[width=0.45\textwidth]{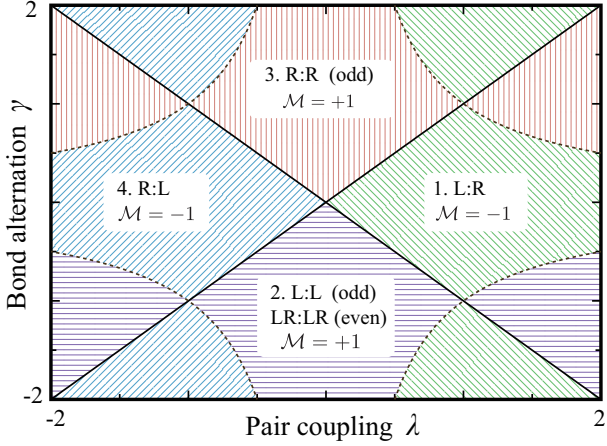}
\caption{Phase diagram of the Kitaev chain with bond alternation. 
The solid line $|\lambda|=|\gamma|$ and dashed line $|\lambda|=|\gamma|^{-1}$ show the phase boundary obtained by the recursive equation and the Pfraffian.
There are four phases that are distinguished by colored regions corresponding to the four cases mentioned in the main text. Green: phase 1 (the case 1), purple: phase 2 (the case 2), red: phase 3 (the case 3), and blue: phase 4 (the case 4). 
The letters such as `L:R' denote the positions of the $a$-type and $b$-type MZMs which is separated by colon, i.e., `L:R' means that the $a$-type MZM is located on the left edge and the $b$-type MZM is on the left edge. 
The terms `odd' (`even') in the bracket denotes that $N$ is odd (even).
The phases without the bracket give the same result , both cases of the odd and even number are the same regardless of even or odd number.
$\mcl{M}$ is the Majorana number obtained by Pfaffian.
\label{fig:pd}}
\end{figure}

Figure~~\ref{fig:pd} shows phase diagram for the $\gamma$-$\lambda$ plane, where each phase represents the corresponding case mentioned above as indicated by the number in each phase.
We note that this phase diagram is an extended version of Ref.~\cite{Wakatsuki14}, i.e., we determine phases in the region of $|\lambda|>1$ and/or $|\gamma|>1$ by extending the region near $|\lambda|=|\gamma|=1$~\cite{note1}.

In Fig.~\ref{fig:pd}, we find that the phase boundaries are composed by not only $|\lambda|=|\gamma|$ but $|\lambda|=|\gamma|^{-1}$.
The phase boundary $|\lambda|=|\gamma|^{-1}$ is understood by a global U(1) rotation of spins belonging to a sublattice in the spin model (\ref{eq:spiham}).
The reason is as follows.
For example, if we consider the region of $\lambda>1$, the local U(1) rotation around $x$ axis, $\mcl{R}_j^x=\exp\lt[\im \pi\sigma_j^x/2\rt]$, changes the sign of $y$ ($z$) component of spin: $-\sigma_j^{y\,(z)}=\mcl{R}_j^x\sigma_j^{y\,(z)}(\mcl{R}_j^x)^\dagger$. 
Therefore, the global U(1) rotation of odd sites $\mcl{R}^x=\prod_{i} \mcl{R}_{2i-1}^x$ ($i=1,2,\cdots\leq(N+1)/2$) changes only the sign of $\sigma_j^y \sigma_{j+1}^y$ terms in (\ref{eq:spiham}), which means a duality between two points locating at $\lambda>1$ and $0<\lambda<1$ in the phase diagram.
Similarly, the global rotation of odd bonds $\mcl{R}^z=\prod_{j} \mcl{R}_{j}^z$ ($j=1,2,5,6,9,10,\cdots\leq N$) maps the region of $\gamma>1$ to that of $0<\gamma<1$.

Remarkably, we can see a reentrant phenomenon as the pairing potential $\lambda$ (the bond alternation $\gamma$) increases with fixed $\gamma$ ($\lambda$): e.g., if we change the pairing potential $\lambda$ from -2.0 to 2.0 with fixed the bond alternation $\gamma=0.75$, we come across four phase transitions and two reentrances to the phase 3.
The phase 1 and the phase 4 give $\mcl{M}=-1$ corresponding to the MZM phase where the $a$-type MZM is located on one of the edges  while the $b$-type MZM is on the other edge.
The phase 2 and the phase 3 give $\mcl{M}=1$, which corresponds to the $Z_2$ topologically ordered phase.
If $N$ is odd, MZM always exists because the boundary constraints are imposed on even-site modes.
However, the phase boundaries are the same as the  case of the even number of $N$.

\subsection{Fidelity}
To confirm the phase transition, we investigate fidelity defined as
\begin{equation}
\mrm{Fd}(\lambda,\gamma;\delta\lambda,\delta\gamma)=\sum_j |u_{\mrm{k}_0,j}(\lambda,\gamma)u_{\mrm{k}_0,j}(\lambda+\delta\lambda,\gamma+\delta\gamma)|,
\end{equation}
where $u_{\mrm{k}_0,j}(\lambda,\gamma)$ satisfies the recursive equation (\ref{eq:uv}) for a given bond alternation $\gamma$ and pairing interaction $\lambda$, corresponding to the $a$-type MZM.
Here, we consider only finite-size systems with the odd number of sites, because there is no solution for MZMs in finite-size systems with even number.
Since $u_{\mrm{k}_0,j}$ is considered as a wavevector of superposing real-space Majorana fermions, the fidelity is expressed in the limit $\delta\lambda\to0$ and $\delta\gamma\to0$ as
\begin{equation}
\mrm{Fd}(\lambda,\gamma;\delta\lambda,\delta\gamma)=\lt[1-\frac{1}{2}\lt(\delta\lambda\frac{\pder}{\pder \lambda}+\delta\gamma\frac{\pder}{\pder \gamma}\rt)\rt] \sum_j |u_{\mrm{k}_0,j}|^2.
\end{equation}
If there is a MZM satisfying $\sum_j |u_{\mrm{k}_0,j}|^2=1$, the fidelity goes to the unity.
However, at the point where the MZM disappears, fidelity cannot be defined and shows discontinuity.
Furthermore, for small but finite value of $\delta\lambda$ and $\delta\gamma$, fidelity sharply drops at the point where the position of MZM changes from one edge to the other edge, because there is only negligibly small overlap between the two MZMs at the left and right edges.

\begin{figure}[t]
\includegraphics[width=0.45\textwidth]{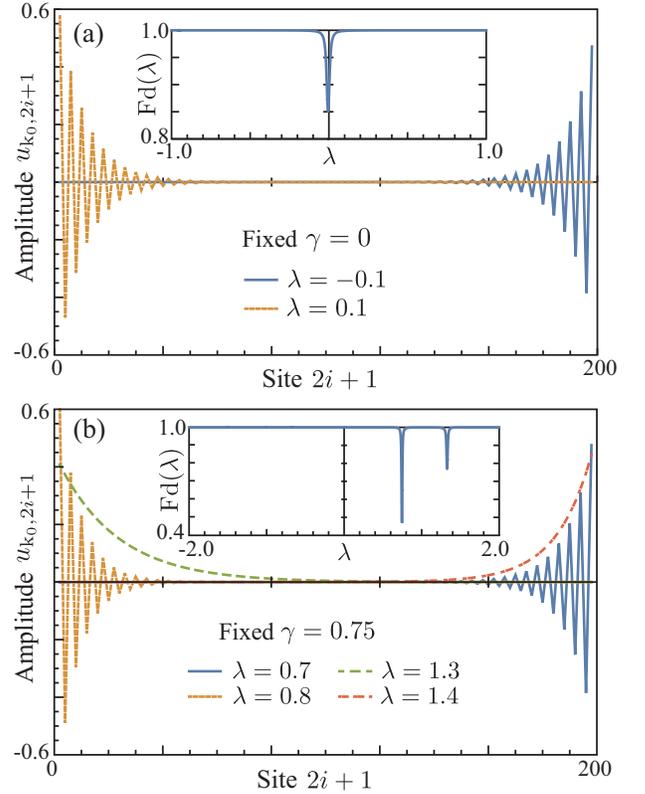}
\caption{Amplitude of Majorana edge mode $u_{\mrm{k}_0,2i+1}$ ($i=0,1,\cdots,99$) for fixed (a) $\gamma=0$ and (b) $\gamma=0.75$ in a $N=199$ sites system.
The inset shows fidelity $\mrm{Fd}(\lambda,\gamma;\delta\lambda=0.01,\delta\gamma=0)$.
\label{fig:fid}}
\end{figure} 

Figure~\ref{fig:fid} shows $u_{\mrm{k}_0,j}$ as a function $j$ in the $N=199$ chain.
At $\delta\gamma=0$, $u_{\mrm{k}_0,j}$ exhibits large amplitude near the left (right) edge for $\lambda=0.1$ ($\lambda=-0.1$) as shown in Fig.~\ref{fig:fid}(a).
This is consistent with the behavior of $u_{\mrm{k}_0,j}$ expected from the phase diagram (Fig.~\ref{fig:pd}).
Changing $\lambda$ from 0.1 to $-$0.1, we have switching of the $a$-type MZM from left to right edge at $\lambda=0$ where the drop of fidelity appears.
This is clearly seen in the inset of Fig.~\ref{fig:fid}(a).
In Fig.~\ref{fig:fid}(b), $u_{\mrm{k}_0,j}$ at $\gamma=0.75$ is shown.
Since there are two boundaries at $\lambda=0.75$ and $\lambda=4/3$ along the $\gamma=0.75$ line in Fig.~\ref{fig:pd}, main amplitude of $u_{\mrm{k}_0,j}$ in Fig.~\ref{fig:fid}(b) is located at the left (right) region for $\lambda=0.7$ and $\lambda=1.3$ ($\lambda=0.8$ and $\lambda=1.4$).
The drop of fidelity at $\lambda=\gamma=0.75$ in the inset of Fig.~\ref{fig:fid}(b) is similar to that at $\lambda=\gamma=0$ because the two points are on the same boundary.
On the other hand, the drop at $\lambda=4/3$ comes from different phase boundary in Fig.~\ref{fig:pd}. 

\subsection{Pfaffian}
To clarify whether the phase transition does corresponds to a topological transition, we examine the Majorana number when the twisted boundary condition is imposed.
We assume that the number of sites is even, because bond alternation cannot be defined consistently in a ring with the odd number.

The twisted boundary condition is introduced by adding boundary Hamiltonian
\begin{equation}
\mcl{H}_{\mrm{bound.}} = t(1-\gamma\ex^{\im\pi N}) (e^{\i \phi} c_N^\dagger c_{1} + e^{\i \phi} \lambda c_N^\dagger c_{1}^\dagger +\mrm{H.c.}), 
\end{equation}
where $\phi$ is the phase of twisted boundary.
The full Hamiltonian is rewritten by
\begin{equation}
\mcl{H}_{\mrm{bulk}}+\mcl{H}_{\mrm{bound.}}=\frac{\im t}{4}\sum_{j,l} d_j h_{j,k}(\phi) d_k,
\end{equation}
where $d_{2j-1}=a_j$, $d_{2j}=b_j$.
By using anti-symmetric matrix $\mbf{h}(\phi)=\{h_{j,k}(\phi)\}=-\{h_{k,j}(\phi)\}$ and Pfaffian of $\mbf{h}(\phi)$, the Majorana number is given by $\mcl{M}=\mrm{sgn}(\mrm{Pf}[\mbf{h}(0)]\times \mrm{Pf}[\mbf{h}(\pi)])$~\cite{Miao16,Kawabata17}.
The Pfaffian of anti-symmetric $2n$-by-$2n$ matrix $\mbf{A}$ generally reads
\begin{equation}
\mrm{Pf}[\mbf{A}]=\sum_{k=2}^{2n} (-1)^k A_{1,k} \mrm{Pf}[\mbf{A}^{\{1,k\}}],
\end{equation}
where $\mbf{A}^{\{1,k\}}$ is the $2(n-1)$-by-$2(n-1)$ minor matrix of $\mbf{A}$ that is obtained by deleting the 1st and $k$th rows as well as those columns.
After performing a recursive procedure (see Appendix A), we obtain the Pfraffian of the coupling matrix $\mbf{h}$ as follows,
\begin{widetext}
\begin{equation}
\mrm{Pf}[\mbf{h}(\phi)] = \lt[(1+\gamma)^N +(1-\gamma)^N\rt](\lambda^2-1)^{N/2} + (1-\gamma^2)^{N/2} [(1-\lambda)^N+(1+\lambda)^N] \cos\phi.
\end{equation}  
\end{widetext}
We thus obtain the Majorana number
\begin{equation}
\mcl{M}=\mrm{sgn}\lt[\lt(1+\Gamma_1^N\rt)^2 \lt(-\Lambda\rt)^{N} -  \lt(-\Gamma_1\rt)^{N} \lt(1+\Lambda^N\rt)^2\rt].
\end{equation}
We can easily verify that the boundary where $\mcl{M}$ changes from $1$ (trivial) to $-1$ (non-trivial) corresponds to the phase boundary in Fig.~\ref{fig:pd}.

\section{Bulk Properties of Spinless Fermions}
Next, we discuss bulk properties in our model: dispersion relations of the quasi-particle and a string order of the Haldane state.

\subsection{Disperison relation and winding number}
The momentum $k$ representation of the Hamiltonian is given by
\begin{equation}
\mcl{H}_{\mrm{bulk}}+\mcl{H}_{\mrm{bound.}}|_{\phi=0}=\frac{t}{2}\sum_k \bm{c}_k^\dagger \mbf{H}_k \bm{c}_k+\mrm{const.}
\end{equation}
with
\begin{equation}
\mbf{H}_k=
\begin{pmatrix}
\cos k & \im \lambda \sin k & \im \gamma\sin k & \lambda\gamma \cos k\\
-\im \lambda \sin k &-\cos k & -\lambda\gamma \cos k & -\im \gamma\sin k\\
-\im \gamma\sin k & -\lambda\gamma \cos k & -\cos k & -\im \lambda \sin k\\
\lambda\gamma \cos k & \im \gamma\sin k &  \im \lambda \sin k & \cos k
\end{pmatrix}
\ ,
\end{equation}
where we use the Nambu representation in the fermionic vector space $\bm{c}_k=(c_k,c_{-k}^\dagger,c_{k+\pi},c_{-k-\pi}^\dagger)^{\mrm{T}}$.
The matrix $\mbf{H}_k$ is rewritten as a linear combination of direct products of the Pauli matrices $\bm{\sigma}$,
\begin{equation}
\mbf{H}_k=(\sigma_z\otimes\sigma_z-\lambda\gamma\sigma_y\otimes\sigma_y)\cos k - (\lambda \sigma_z\otimes\sigma_y+\gamma \sigma_y\otimes\sigma_z)\sin k.
\end{equation}
In this formalism, we can obtain a block-diagonal matrix as follows,
\begin{eqnarray}
\mbf{U}_k\mbf{H}_k\mbf{U}_k^\dagger&=&\lt[-(1+\lambda\gamma)\cos k\,\sigma_z+(\lambda+\gamma)\sin k \,\sigma_y\rt] \nonumber \\
& &\oplus\lt[(1-\lambda\gamma)\cos k\,\sigma_z-(\lambda-\gamma)\sin k \,\sigma_y\rt] \label{eq:block}
\end{eqnarray}
with a unitary matrix
\begin{equation}
\mbf{U}_k=\exp\lt[\im\frac{\pi}{4}\sigma_y\otimes(2-\sigma_x)\rt].
\end{equation}
Therefore, we find the eigenvalues $\epsilon_1^\pm$ and $\epsilon_2^\pm$ by diagonalizing the submatrices:
\begin{equation}
\epsilon_1^\pm=\pm\sqrt{\frac{(1+\lambda\gamma)^2+(\lambda+\gamma)^2+(\lambda^2-1)(\gamma^2-1)\cos(2k)}{2}}
\end{equation}
and 
\begin{equation}
\epsilon_2^\pm=\pm\sqrt{\frac{(1-\lambda\gamma)^2+(\lambda-\gamma)^2+(\lambda^2-1)(\gamma^2-1)\cos(2k)}{2}}.  
\end{equation}
If we consider positive values of $\lambda$ and $\gamma$, it is enough to discuss only $\epsilon_2^\pm$ to determine the bulk gap because $|\epsilon_2^\pm|<|\epsilon_1^\pm|$.
Therefore, the bulk gap closes at $k=\frac{\pi}{2}$ for $\lambda=\gamma$ and at $k=0$ for $\lambda=\gamma^{-1}$.
This implies that the two different boundaries in Fig.~\ref{fig:pd} have different characteristic even in the momentum space, where the bulk gap closes at the different momentum.

Furthermore, we can define a winding number in Eq.~(\ref{eq:block}) with mapping the Pauli matrices to the unit vectors $(\sigma_y,\sigma_z)\to(\hat{\bm{y}},\hat{\bm{z}})$~\cite{Read00}. 
This mapping gives an $\mbb{R}^2\oplus\mbb{R}^2$ representation of the block-diagonalized Hamiltonian (\ref{eq:block}), that is, extended Anderson pseudo-vectors~\cite{Read00,Anderson58} obtained as $\bm{v_1}\oplus\bm{v_2}$ with $\bm{v_1}=-(1+\lambda\gamma)\cos k\,\hat{\bm{z}}+(\lambda+\gamma)\sin k \,\hat{\bm{y}}$ and $\bm{v}_2=(1-\lambda\gamma)\cos k\,\hat{\bm{z}}-(\lambda-\gamma)\sin k \,\hat{\bm{y}}$.
Thus, the winding number is defined by
\begin{equation}
n_{\mrm{w}}=\frac{1}{2}\sum_{i=1,2}\oint_k \frac{\der\,(\mrm{arg}\,\bm{v}_i)}{2\pi},
\end{equation}
where the integrated region is given by $k\in[0,2 \pi]$, which is consisted with the result in Ref.~\cite{Wakatsuki14}.
The Majorana number is also given by the winding number $\mcl{M}=\cos\lt(\pi n_{\mrm{w}}\rt)$, which is consistent with the Majorana number obtained by the Pfaffian.

\subsection{String order of Haldane state}
\begin{figure}[t]
\includegraphics[width=0.45\textwidth]{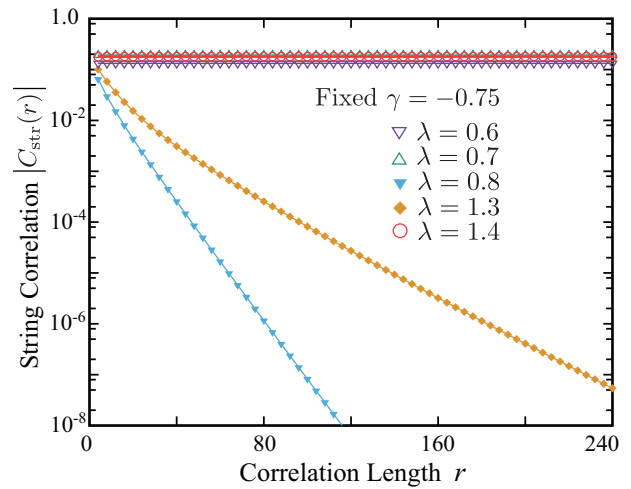}
\caption{ Correlation function of the string order for various $\lambda$ with fixed $\gamma=-0.75$.
\label{fig:cs}}
\end{figure}

Finally, we investigate a string order of the Haldane state as a topological order parameter in the bulk.
As a natural extension of the string order in an $S=1$ Haldane chain, we examine a correlation function of the string order as follows, 
\begin{equation}
C_{\mrm{str}}(r) = \frac{1}{4} \lag(\sigma_{L_r}^z+\sigma_{L_r+1}^z)\lt(\prod_{j=L_r+2}^{R_r-1} \sigma_{j}^z\rt)(\sigma_{R_r}^z+\sigma_{R_r+1}^z)\rag,
\end{equation}
where the left (right) site is defined by $L_r=N/2-r/2+\theta(\gamma)$ ($R_r=N/2+r/2+\theta(\gamma)$) with the Heviside step function $\theta(x)$.
This correlation function is easily calculated for two cases: (i) $\lambda=1, \gamma=0$ and (ii) $\lambda=0, \gamma=-1$.
The case (i) and the case (ii) correspond to the topological state ($\mcl{M}=-1$) with a MZM and the topologically-ordered state with a trivial Majorana number $\mcl{M}=1$, respectively.
The ground state of the case (i) is given by a MPS representation,
\begin{equation}
|\Psi_{\mrm{(i)}}^\pm\rag = 2^{-N/2} \lt(\prod_{j=1}^N\ex^{\pm c_j^\dagger}\rt)|0\rag,
\end{equation}
where $|0\rag$ denotes the vacuum, i.e., $c_j|0\rag=0$ for an arbitrary $j$~\cite{Katsura15}.
The relation $\lag 0|\ex^{\pm c_j} \sigma_j \ex^{\pm c_j^\dagger} |0\rag= \lag 0|\lt[n_j\pm(c_j+c_j^\dagger)\rt]|0\rag=0$ gives the trivial correlation function $C_{\mrm{str}}(r)=0$.
On the other hand, the groud state of the case (ii) is a direct product of singlet states as follows,
\begin{equation}
|\Psi_{\mrm{(ii)}}\rag=2^{-N/4} \lt[\prod_{i=1}^{N/2} \bm{v}_{2i-1}^{\mrm{(L)}} \cdot \bm{v}_{2i}^{\mrm{(R)}} \rt] |0\rag,
\end{equation}
where the left and right vector operator are defined by $\bm{v}_{j}^{\mrm{(L)}}=(c_j^\dagger,1)$ and $\bm{v}_{j}^{\mrm{(R)}}=(1,-c_j^\dagger)$, respectively.
In this case, the correlation function has a non-zero constant value $C_{\mrm{str}}(r)=1/4$.
Figure~\ref{fig:cs} shows the correlation function of the string order $C_{\mrm{str}}(r)$ with fixed $\gamma=-0.75$.
This is numerically obtained by variational MPS calculation for an $N=512$ system~\cite{schollwock11,note2}.
In Fig.~\ref{fig:cs}, we can see that the correlation function converses to finite value with increasing the length $r$ in the $\mcl{M}=1$ phase, whereas it exponentially decreses in the topological phase with an MZM ($\mcl{M}=-1$).
Therefore, the $\mcl{M}=1$ phase corresponds to the Haldane state, where non-zero string order parameter emerges in the bulk.

\section{Summary}
We theoretically study the effects of bond alternation on the Kitaev chain, as an extention of preceding work~\cite{Wakatsuki14}.
Three analytical approaches, the recursive equation of MZM, the fidelity of MZM amplitude, and the Pfaffian of coupling matrix, are used to examine phase transition between the MZM phase and the Z$_2$ topologically ordered phase.
In the extended phase diagram, it is found that there are two phase boundaries with a reentrant phenomenon, where the bulk gap closes at different momenta. 
In addition, we find that the phase transition is caused by switching the edge position of MZMs, and we can distinguish the two phases transition with the Majorana number $\mcl{M}$ obtained by the Pfaffian.
Several preceding studies have reported that the MZM is robust against perturbations such as a repulsive interaction and disorders~\cite{Katsura15,Sela11,Hassler12,Niu12,Brouwer11,Lobos12,DeGottardi13}, and thus the Kitaev model is expected as a quantum memory~\cite{Mazza13,Ivanov01}, e.g., in a quantum nano wire~\cite{Mourik12,Rokhinson12}.
Consequently, our study not only provides a simple model bridging between the bulk and edge topologies, but also indicates the possibility of the reentrant topological phase transition in real systems~\cite{note1,Tezuka13}.

{\it Note added.} While this paper was in review, an article [M. Ezawa, Phys. Rev. B {\bf 96}, 121105(R) (2017).] discussing effects of interaction in our model was published.

\begin{acknowledgements}
We would like to thank S. A. Jafari and H. Katsura for fruitful discussions and useful informations. 
\end{acknowledgements}

\appendix*
\section{Calculation of the Pfaffian}
In this section, we perform the calculation of $\mrm{Pf}[\mbf{h}(\phi)]$ where the $2N$-by-$2N$ matrix $\mbf{h}(\phi)$ is anti-symmetric $h_{j,k}(\phi)=-h_{k,j}(\phi)$.
The elements of $\mbf{h}(\phi)$ in our model are given by,
\begin{equation}
h_{2j-1,2j+2}=\gamma_j(1-\lambda), \hspc h_{2j,2j+1}=-\gamma_j(1+\lambda)
\end{equation}
for $j=1,\cdots,N-1$, and at the boundary
\begin{align}
&h_{1,2N-1}=-\gamma_N(1+\lambda)\sin\phi, \hspc h_{2,2N}=-\gamma_N(1-\lambda)\sin\phi,\\
&h_{2,2N-1}=-\gamma_N(1-\lambda)\cos\phi, \hspc h_{1,2N}=\gamma_N(1+\lambda)\cos\phi,
\end{align}
where $\gamma_j=1-\gamma\ex^{\im\pi j}$.
Next, we consider the following relation
\begin{equation}
\mrm{Pf}[\mbf{A}]=\sum_{k=2}^{2n} (-1)^k A_{1,k} \mrm{Pf}[\mbf{A}^{\{1,k\}}] 
\end{equation}
where $\mbf{A}^{\{1,k\}}$ is the $2(n-1)$-by-$2(n-1)$ minor matrix of $\mbf{A}$ that is obtained by deleteing the 1st and $k$th rows, and those columns.
\begin{widetext}
The first step of the expansion reads,
\begin{equation}
\mrm{Pf}[\mbf{h}]= \gamma_1(1-\lambda) \mrm{Pf}[\mbf{h}^{\{1,4\}}] +\gamma_N(1+\lambda)\mrm{Pf}[\mbf{h}^{\{1,2N-1\}}]\sin\phi  +\gamma_N(1+\lambda)\mrm{Pf}[\mbf{h}^{\{1,2N\}}]\cos\phi. 
\end{equation}
In the same manner, we obtain
\begin{align}
\mrm{Pf}[\mbf{h}^{\{1,4\}}]&=-\gamma_1(1+\lambda)\mrm{Pf}[\mbf{h}^{\{1,2,3,4\}}]+\gamma_N(1-\lambda)\mrm{Pf}[\mbf{h}^{\{1,2,4,2N-1\}}]\cos\phi -\gamma_N(1+\lambda)\mrm{Pf}[\mbf{h}^{\{1,2,4,2N\}}]\cos\phi\\
\mrm{Pf}[\mbf{h}^{\{1,2N-1\}}]&=-\gamma_1(1+\lambda)\mrm{Pf}[\mbf{h}^{\{1,2,3,2N-1\}}]-\gamma_N(1-\lambda)\mrm{Pf}[\mbf{h}^{\{1,2,2N-1,2N\}}]\sin\phi \\
 \mrm{Pf}[\mbf{h}^{\{1,2N\}}]&=-\gamma_1(1+\lambda)\mrm{Pf}[\mbf{h}^{\{1,2,3,2N\}}]-\gamma_N(1-\lambda)\mrm{Pf}[\mbf{h}^{\{1,2,2N-1,2N\}}]\cos\phi.
\end{align}
Here, it is noted that $N$ is even and $\gamma_{2j-1}=\gamma_1$, $\gamma_{2j}=\gamma_2$, so that we obtain the following equations:
\begin{align}
&\mrm{Pf}[\mbf{h}^{\{1,2,3,4\}}] = [-\gamma_1^2(1-\lambda^2)]^{N/2-1}, \hspc \mrm{Pf}[\mbf{h}^{\{1,2,2N-1,2N\}}]=[-\gamma_2^2(1-\lambda^2)]^{N/2-1}, \\
&\mrm{Pf}[\mbf{h}^{\{1,2,3,2N\}}]=[\gamma_1\gamma_2(1+\lambda)^2]^{N/2-1}, \hspc \mrm{Pf}[\mbf{h}^{\{1,2,4,2N-1\}}] = [\gamma_1\gamma_2(1-\lambda)^2]^{N/2-1},
\end{align}
and $\mrm{Pf}[\mbf{h}^{\{1,2,3,2N-1\}}]=\mrm{Pf}[\mbf{h}^{\{1,2,4,2N\}}]=0$.
We use these equations in the expansion of the Pfaffian,
\begin{align}
\mrm{Pf}[\mbf{h}^{\{1,4\}}]&= -\gamma_1(1+\lambda) [-\gamma_1^2(1-\lambda^2)]^{N/2-1} + \bar{\gamma_1}(1-\lambda)[\gamma_1\gamma_2(1-\lambda)^2]^{N/2-1}\cos\phi,\\
\mrm{Pf}[\mbf{h}^{\{1,2N-1\}}]&= -\gamma_2(1-\lambda) [-\gamma_2^2(1-\lambda^2)]^{N/2-1}\sin\phi, \\
\mrm{Pf}[\mbf{h}^{\{1,2N\}}] &= -\gamma_1(1+\lambda) [\gamma_1\gamma_2(1+\lambda)^2]^{N/2-1} -\gamma_2(1-\lambda)[-\gamma_2^2(1-\lambda^2)]^{N/2-1}\cos\phi.
\end{align}
We finally obtain
\begin{align}
\mrm{Pf}[h]= \lt[(1+\gamma)^N +(1-\gamma)^N\rt](\lambda^2-1)^{N/2} + (1-\gamma^2)^{N/2} [(1-\lambda)^N+(1+\lambda)^N] \cos\phi.
\end{align}
\end{widetext}


\begin{thebibliography}{00}
\bibitem{note1} The reason why we extend region of the phase diagram comes from the following consideration. In the dimerized chain, the magnitude of two kinds of bond interactions is originally different, but it may be possible to change the sign of one of the two interactions by applying, for example, local pressure to the corresponding bond, modifying bond angle and length contributing to ferromagnetic/antiferromagnetic interactions. This will bring the extended region of $|\lambda|>1$ and/or $|\gamma|>1$. In addition, we expect to observe the reentrant phenomenon by applying local pressure if the effective parameters of a real material are located around $|\lambda|=|\gamma|=1$.
\bibitem{note2} In this calculation, the number of kept states is 200, and the truncation error is less than $10^{-12}$.
\end{thebibliography}
\end{document}